\UseRawInputEncoding
\documentclass[preprint,12pt    ]{elsarticle}
\usepackage{graphicx}
\usepackage{float}
\usepackage{latexsym,amsbsy,epsfig,fancybox}
\usepackage{amstext}
\usepackage{subfigure}
\usepackage{amsfonts,textgreek}
\usepackage{mathrsfs}
\usepackage{mathtools}
\usepackage{tikz}
\usepackage{pstricks}
\usepackage{array}
\usepackage{amssymb}
\usepackage{multirow}
\usepackage{booktabs}
\usepackage{verbatim}
\usepackage{natbib}
\biboptions{sort&compress}
\usepackage{url}
\usepackage{xr}
\usepackage[margin=3cm]{geometry}
\usepackage{soul}
\usepackage[colorlinks=true,citecolor=blue]{hyperref}

\newcommand{\ac}[1]{\textcolor{red}{add citation}}

\newcommand{\af}[1]{\textcolor{red}{add figure}}
\newcommand{\ar}[1]{\textcolor{red}{add crossref}}
\tikzstyle{nicebox}=[draw=black!100, fill=white!10, rectangle, inner sep=4pt, inner ysep=16pt]
\tikzstyle{niceboxtitle}=[draw=black!100, fill=white, text=black, rectangle]

\usetikzlibrary{arrows,decorations.markings}
\newcommand{\change}[1]{{#1}} 
\newcommand{\response}[1]{\textcolor{black}{#1}}
\journal{npj Computational Materials}
\begin{document}
\begin{frontmatter}

\title{Why is the strength of a\response{n elastomeric} polymer network so low?}
\author[1]{Shaswat Mohanty}
\author[2]{Jose Blanchet}
\author[3]{Zhigang Suo}
\author[1]{Wei Cai\corref{cor1}}
\ead{caiwei@stanford.edu}
	
\address[1]{Department of Mechanical Engineering, Stanford University, CA 94305-4040, USA}
\address[2]{Department of Management Science and Engineering, Stanford University, CA 94305-4040, USA}
\address[3]{John A. Paulson School of Engineering and Applied Sciences, Harvard University, MA, 02138, USA}

\cortext[cor1]{Corresponding author}

\begin{abstract}
Experiments have long shown that a polymer network of covalent bonds commonly ruptures at a stress that is orders of magnitude lower than the strength of the covalent bonds. Here we investigate this large reduction in strength by coarse-grained molecular dynamics simulations. We show that the network ruptures by sequentially breaking \change{only} a small fraction of bonds, and that each broken bond lies on \change{a path belonging to the left-tail of the ``shortest paths'' distribution}. A shortest path is the path of the fewest bonds that connect two monomers at the opposite ends of the network. As the network is stretched, 
\change{the strands along these left-tail shortest paths straighten and bear high tension set by covalent bonds, while most strands off these paths deform via entropic elasticity}. After a bond on one of the left-tail shortest paths breaks, \change{the load shifts to other left-tail shortest paths and} the process repeats.
As the network is stretched and bonds are broken, the scatter in lengths of the shortest paths first narrows, causing stress to rise, and then broadens, causing stress to decline. This sequential breaking of a small fraction of bonds causes the network to rupture at a stress that is orders of magnitude below the strength of the covalent bonds.
\end{abstract}


\end{frontmatter}
\section{Introduction} 
\label{sec:Intro}

Elastomers are highly stretchable polymer networks. Attention here is focused on an elastomer in which covalent bonds link monomers into polymer strands, and crosslink the polymer strands into a polymer network. Experiments have long shown that the strength of such an elastomer is orders of magnitude lower than the strength of the covalent bonds. A representative value of the strength of a covalent bond is on the order of $10$ GPa~\citep{yang2019polyacrylamide}, whereas a representative value of the experimentally measured strength of an elastomer is on the order of $10$ MPa~\citep{helaly2011effect,wang2023research}. What causes this three orders of magnitude reduction in strength?

This enormous reduction in strength is also commonly reported in other elastic materials, such as glass and ceramics~\citep{clarke1987fracture, shand1965strength}. Since the work of Griffith~\citep{caceres1995deformation}, this reduction in strength in glass and ceramics has been attributed to crack-like flaws, which concentrate stress to \change{the} atomic scale~\citep{yang2019polyacrylamide,berry1961fracture}. Attempts have been made to apply the Griffith theory to account for the large reduction in strength in polymer networks~\citep{kendall1983relation, berry1961fracture}. These attempts, however, have led to problems. In particular, experiments have shown that the strength of a polymer network, such as a polyacrylamide hydrogel, is insensitive to cracks of length up to about $1$ mm~\citep{yang2019polyacrylamide}. 
\response{In this work, we focus on the intrinsic strength of homogeneous polymer networks, specifically addressing why this value is orders of magnitude lower than the strength of individual covalent bonds.}

\begin{figure}[H]
    \centering
    \includegraphics[width=0.48\linewidth]{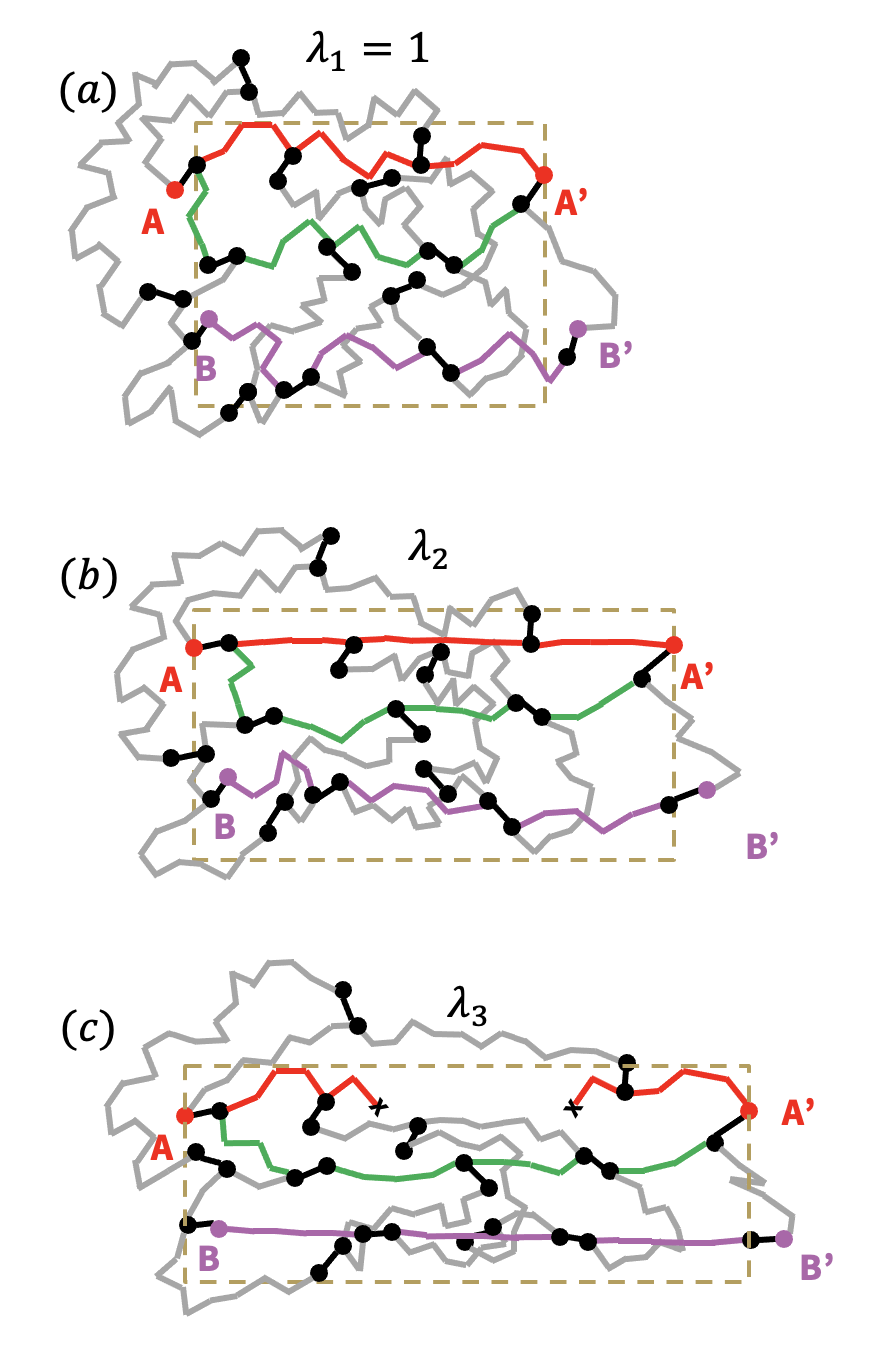}
    \caption{A polymer network ruptures by breaking bonds on a sequence of \change{paths belong to the left-tail of shortest paths distributions}. (a) For any two monomers at the opposite ends of the network, the shortest path is the path of the fewest bonds that connect them, e.g., the shortest path between A and A' (red), and the shortest path between B and B' (purple). Of all the shortest paths, the \change{red and purple paths are the shortest, i.e. on the left-tail of the length distribution}. (b) When the network is stretched to $\lambda_2$, the left-tail shortest paths \change{are} stretched nearly straight, near the breaking point, while most other strands deform by entropic elasticity. (c) At a further stretch $\lambda_3$, a bond breaks on \change{the shortest path A-A'}. 
    \change{The shortest path B-B' becomes even straighter and bears higher load than before, and the process repeats itself.}
    }
    \label{fig:schematic}
\end{figure}

We have previously established the shortest path (SP) between distant crosslinks as a key microstructural feature controlling the macroscopic behavior of the elastomers~\citep{yin2020topological,yin2024network,zhang2024modeling}. Here we hypothesize that the large reduction in strength in a polymer network is due to sequential breaking of a small fraction of bonds, which are not localized around a crack but are distributed throughout the network. We test this hypothesis using coarse-grained molecular dynamics (CGMD) simulations. Our simulations confirm that the network ruptures by sequentially breaking a small fraction of bonds, and each broken bond lies on 
\change{paths belonging to the left tail of the distribution of ``shortest paths''}~\cite{mohanty2024understanding}. Connecting each pair of monomers at the opposite ends of the network is a shortest path
\change{defined as the path with the fewest bonds} (Fig.~\ref{fig:schematic}(a)).
As the network is stretched, the \change{left-tail shortest paths straighten and bear} high tension set by the covalent bonds, while most strands off \change{these paths} deform by entropic elasticity (Fig.~\ref{fig:schematic}(b)). After a bond on \change{a left-tail} shortest path breaks, the \change{load increases further on the remaining left-tail shortest paths and the} process repeats \change{itself} 
(Fig.~\ref{fig:schematic}(c)). 
Because of the statistical scatter in the lengths of the \change{shortest paths}, only an \change{very} small fraction of the paths \change{(i.e. those belonging to the left tail)} bear the high tension approaching the covalent bond strength, while the vast majority of the paths bear low stress. These few high-tension paths contribute negligibly to overall load-bearing, while the majority of low-tension paths bear most of the load. It is this network ``imbalance'' that causes the polymer to rupture at a stress that is orders of magnitude below the strength of the covalent bonds. 
 
A long tradition exists that attributes the large reduction in strength to the scatter of the lengths of polymer strands in a network~\citep{tehrani2017effect,edgecombe1998role}. In an idealized model, a network is represented by parallel strands connecting two rigid and parallel plates~\citep{tao2023effect}. As the two plates are pulled apart, the shortest strand reaches the covalent bond stretch and breaks, while the other strands still carry low stress. For such a model to reduce strength by orders of magnitude, the distribution of lengths of strands needs to be extremely wide. More significantly, this parallel strand model misses a significant aspect of a polymer network. Many strands in a network are extremely short but do not necessarily bear high tension or break early. In a network, strands are connected neither in parallel nor in series. Tension in a strand can be transmitted to the strands connected to it, so that tension usually does not build up to a level sufficient to break a strand. Our earlier work shows that the controlling microstructural parameter for strain-induced damage is the shortest path length connecting far-away monomers~\citep{yin2020topological,yin2024network}, which has motivated related research in polymer rupture analysis~\cite{yu2025shortest}. The relevant shortest path is defined between two monomers separated by a large distance so that upon stretching, their distance vector deforms affinely with the applied stretch. The same cannot be said for the distance vector connecting the two ends of a typical polymer strand.  Hence the concept of the shortest path provides a measure of the connectivity of the network that goes beyond the local measures such as the length of polymer strands, and does not suffer from the same limitations mentioned above. 
\response{Shortest path concepts have been previously used to rationalize failure in highly cross-linked polymer adhesive layers constrained between rigid substrates~\citep{stevens2001interfacial, stevens2001manipulating}. 
However, similar to the parallel-plates model~\citep{tao2023effect} above, the existence of two rigid substrates in this model means that it cannot explain the strength reduction in homogeneous elastomers devoid of any interfaces or defects. More recently, \citet{yu2025shortest} applied shortest-path analysis to thermoset networks and showed that early bond breaking events correlate strongly with the instantaneous minimum shortest paths, building on earlier topological analyses of elastomeric networks~\citep{yin2020topological, yin2024network}. 
Unfortunately, a fundamental gap persists between the brittle fracture observed in thermoset experiments (occurring at small strain) and the ductile failure seen in thermoset simulations (occurring at large strain).
This discrepancy is absent in elastomeric networks -- the primary focus of this study.
%
}

The rest of the paper is structured as follows. The Results Section presents the CGMD simulation results that capture the orders of mismatch between the ideal rupture stress of the cross-link and the peak stress carried by the polymer network. The Discussions Section presents the discussion on the microstructural evolution encoded by the SP distribution between distant cross-links, as the key indicators of material failure. We summarize our findings in the concluding remarks of the Discussions Section. The Methods Section outlines the simulation and analysis methods employed to study the microstructural evolution of the elastomer under load.

\section{Methods} \label{sec:methods}
\subsection{CGMD Model}
In our CGMD simulations, the polymer network is represented by the commonly used bead-spring (Kremer-Grest) model~\citep{kremer1990dynamics}.  We perform the CGMD simulations using LAMMPS~\citep{LAMMPS}.  The simulation cell is subjected to periodic boundary conditions in all three directions and contains $500$ chains, each consisting of $500$ beads. The chains are initialized as self-avoiding random walks. The covalent bonds along a polymer strand are represented by a quartic bond potential and the non-covalent interactions between strands are modeled by a Lennard-Jones (LJ) potential \response{following prior coarse-grained network rupture studies (e.g., Kremer--Grest-type networks and related extensions)}~\citep{yin2020topological,yin2024network,kremer1990dynamics,stevens2001manipulating,stevens2001interfacial}. The initial configuration is first equilibrated to form a polymer melt~\citep{sliozberg2012fast}. The crosslinks, also represented by quartic bonds, are then added between randomly chosen pairs of neighboring beads~\citep{zhang2024modeling}. The CGMD simulation cell is a cube with a side length of $\sim55.4$ nm. We use a total of $8000$ cross-links to have a cross-link density of $\sim 9.66 \times 10^{-5}$ mol/cm$^3$. The effective diameter of the repeat unit of the natural rubber backbone is $\sigma = 4.89$~\AA~\citep{uddin2016multiscale} \response{and the depth of the energy well for pairwise interactions is $\epsilon=1.59$ kJ/mol~\cite{chaikumpollert2012mechanical}}. 
\response{The quartic potential~\citep{ge2013molecular}, $U_{\rm Q}$ is described as
\begin{equation}
    U_{\rm Q}(r) = 
    \begin{cases}
    K(r-R_c)^2(r-R_c-B_1)(r-R_c-B_2)+U_0, \text{  for }r\leq R_c,\\
    0, \text{  for }r>R_c.
    \end{cases}
\end{equation}
Here $K=1200\,\epsilon/\sigma^4$ is the bond stiffness, $B_1=-0.55\,\sigma$, $B_2=0.95\,\sigma$, $U_0=34.6878\,\epsilon$, and $R_c=1.3\,\sigma=6.357\,$\AA ~is the cutoff distance beyond which the quartic bond is considered broken.}

The resulting polymer network is then further equilibrated under the NPT ensemble at $T = 300$ K and at zero pressure for $10\, \mu$s, followed by additional equilibration under the NVT ensemble at $T = 300$ K for another $10\, \mu$s. The equilibrated polymer network is then stretched in one direction up to ten times its original length at a rate of $8.7 \times 10^5$ s$^{-1}$ at $T = 300$ K. The lateral directions contract to preserve the volume during stretching, \response{i.e., isochoric deformation (volume-preserving) is enforced to maintain $\lambda_x\lambda_y\lambda_z=1$ by setting $\lambda_y = \lambda_z = \lambda_x^{-1/2}$.} In this stage of the simulation, covalent bonds can break but cannot form. The positions of all the beads and covalent bond connectivities are stored at an interval of $0.001$ in stretch \response{for subsequent analysis}.

\subsection{Shortest Path Analysis} \label{subsec:sp_analysis}
We convert the polymer network from the CGMD representation into a graph where only the crosslinked beads are kept as nodes. An edge exists between two nodes if the corresponding beads are connected through either a crosslink or a polymer strand. The weight of the edge is the number of bonds on that polymer strand, or $1$ when the two beads are connected by a cross-link. The shortest path \response{(SP)} is computed between every pair of nodes that are separated in the direction of stretch by the simulation cell size ~\citep{zhang2024modeling,yin2024network}, using Dijkstra's algorithm~\citep{dijkstra2022note}. \response{The SP distribution is the distribution of shortest-path lengths (measured as number of bonds) over all node pairs described above.}

\section{Results} \label{sec:results}
\subsection{Stress-stretch Response}
\begin{figure}[H]
    \centering
    \begin{minipage}[t]{0.36\textwidth} 
        \centering
        \subfigure[]{\includegraphics[width=0.9\textwidth]{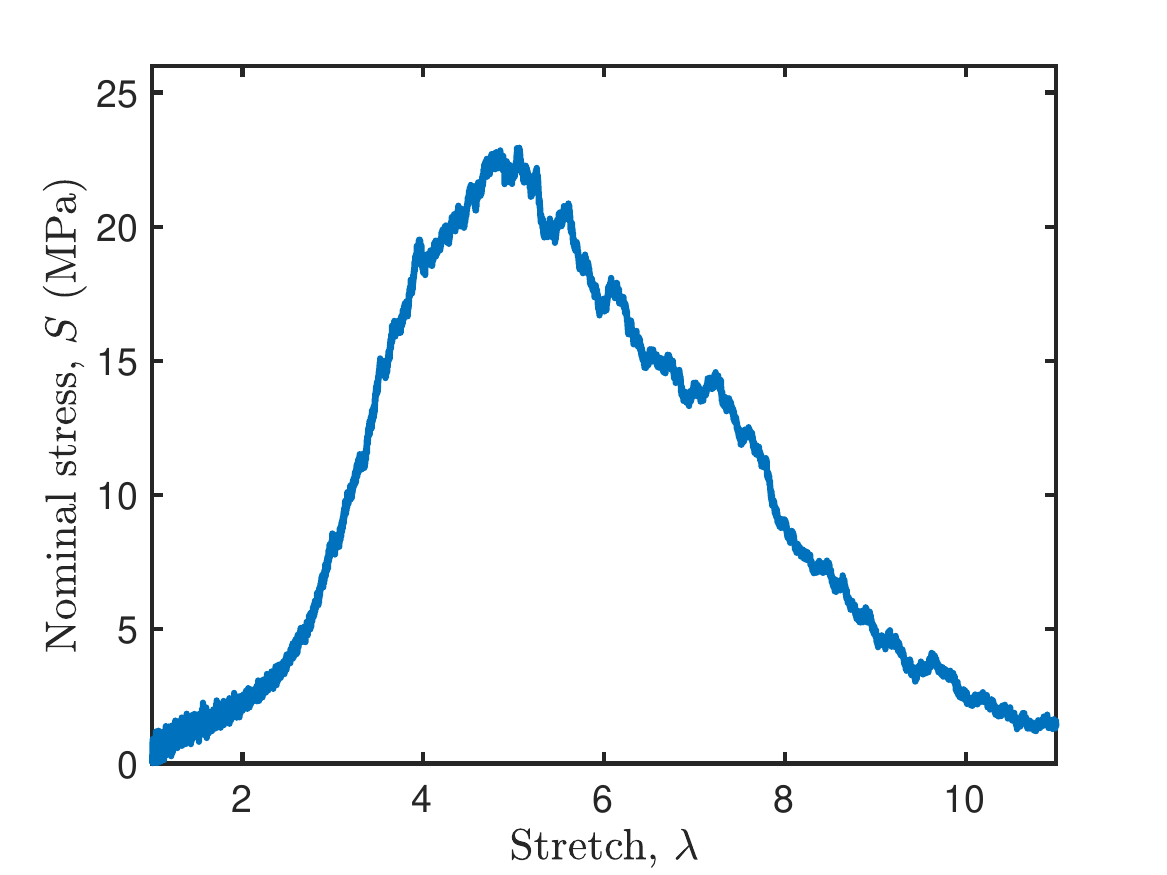}\label{fig:reg_hist_a}}\\ 
        \subfigure[]{\includegraphics[width=0.9\textwidth]{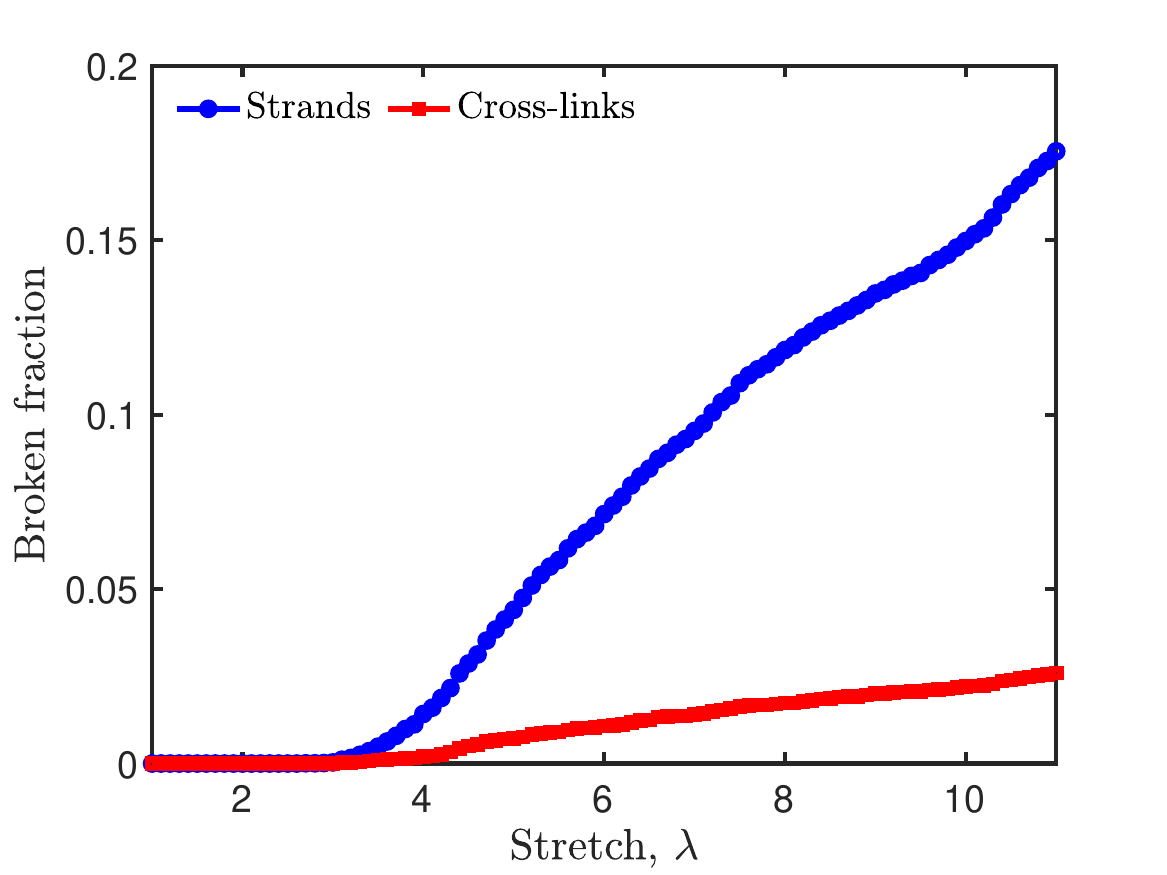}\label{fig:reg_hist_b}} 
    \end{minipage}%
    \hfill
    \begin{minipage}[t]{0.62\textwidth}
        \centering
        \subfigure[]{\includegraphics[height=0.39\textwidth,width=0.98\textwidth]{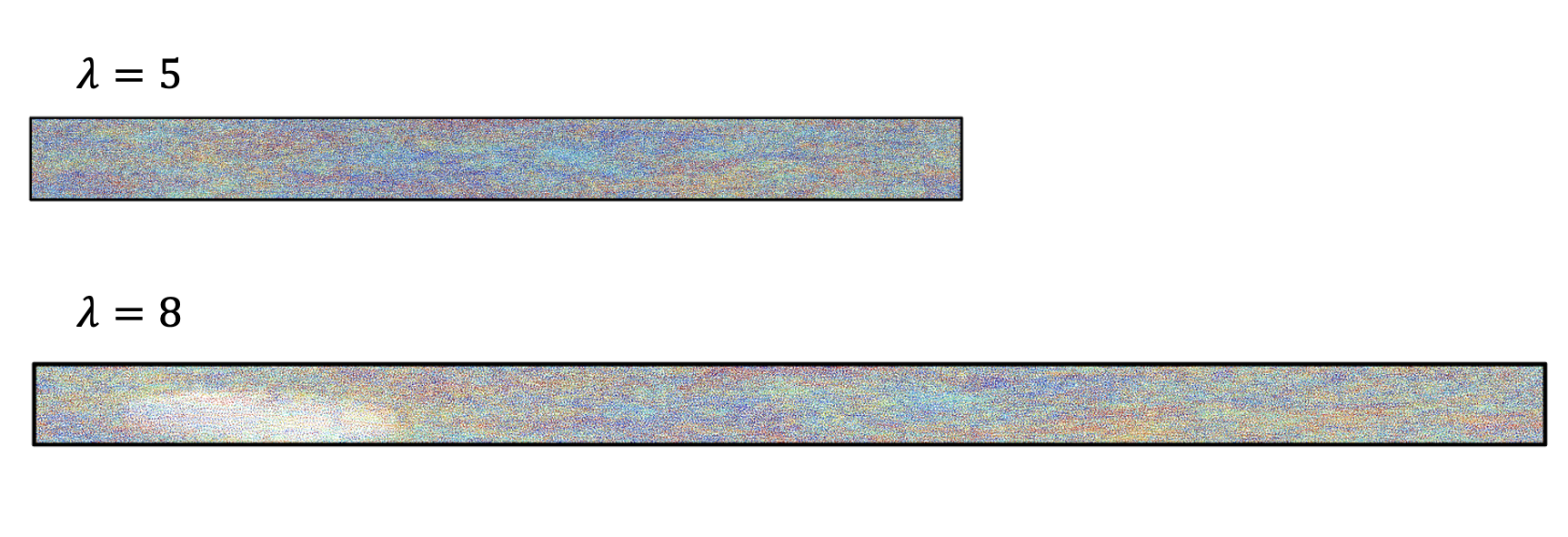}\label{fig:reg_hist_c}}   
        \subfigure[]{\includegraphics[height=0.395\textwidth,width=0.49\textwidth]{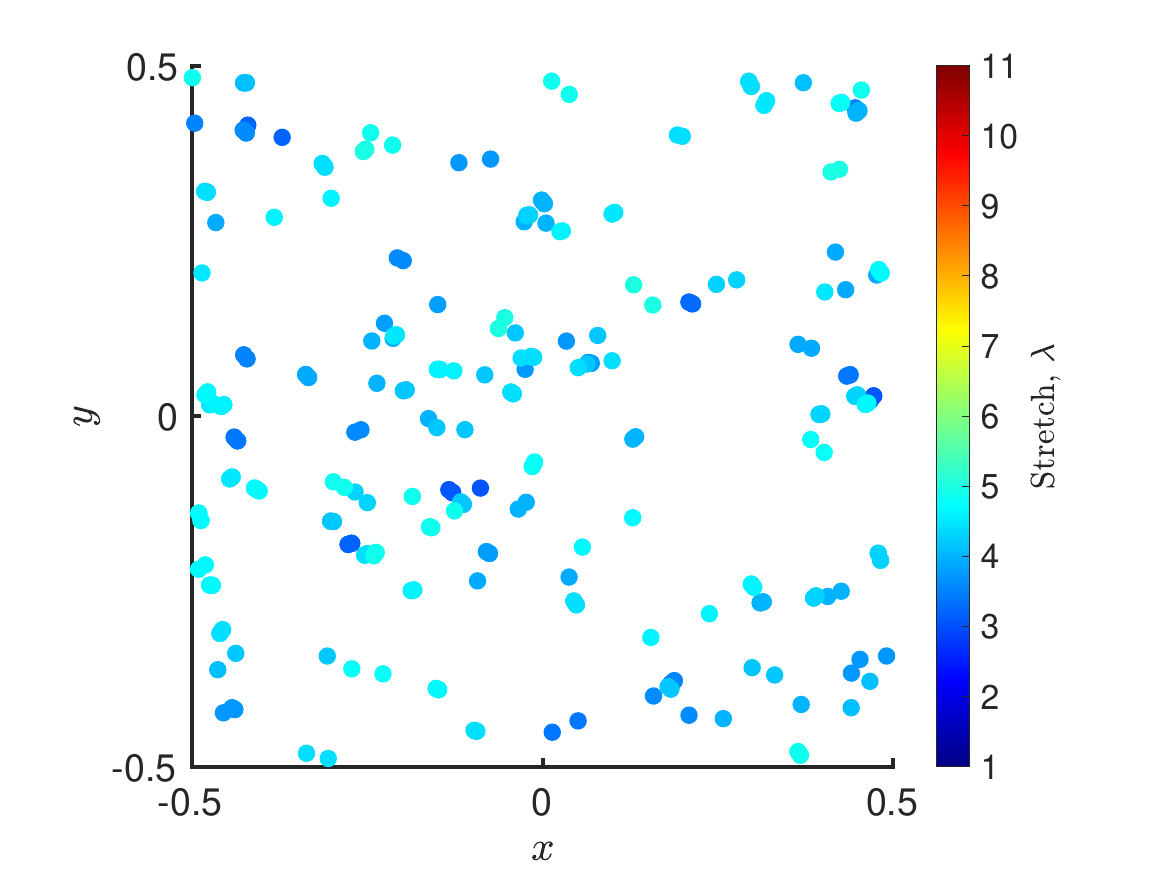}\label{fig:reg_hist_d}}
        \subfigure[]{\includegraphics[height=0.395\textwidth,width=0.49\textwidth]{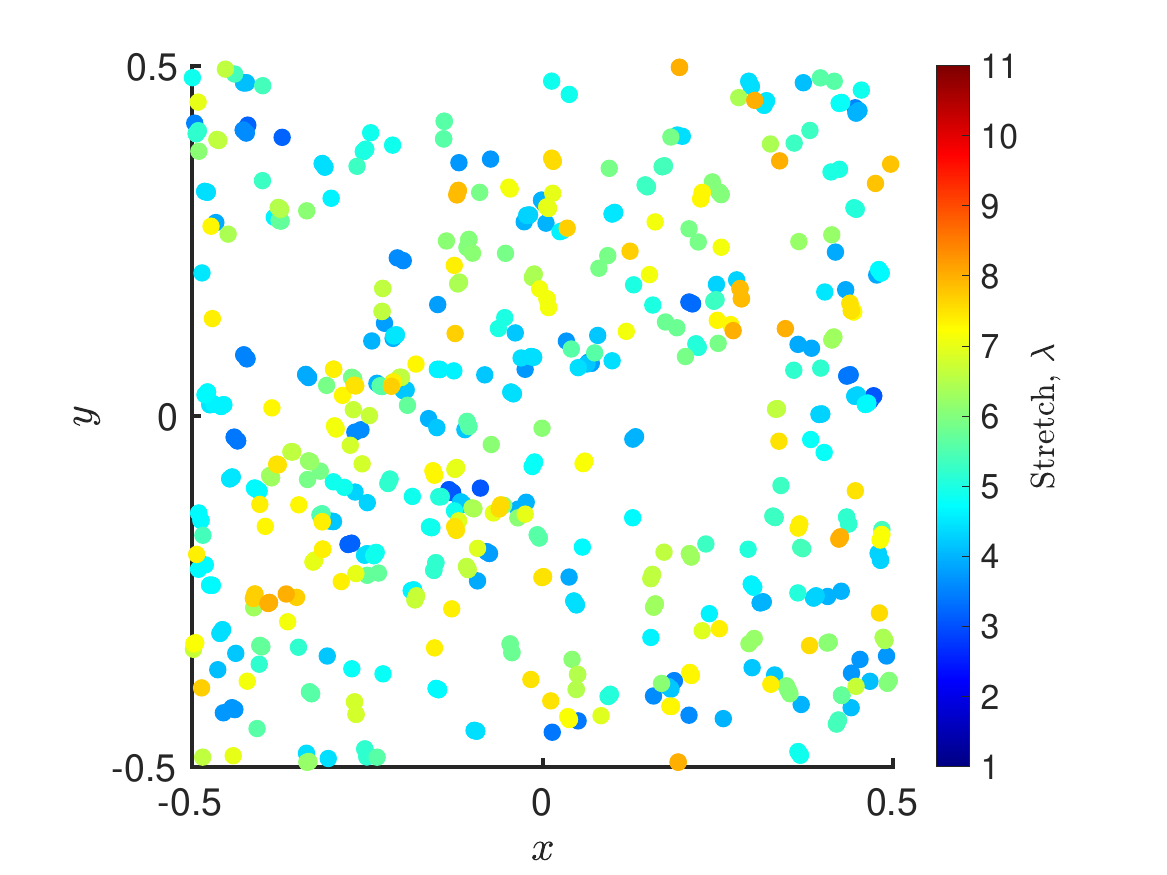}\label{fig:reg_hist_e}}
    \end{minipage} 
    \caption{The network ruptures by breaking a small fraction of strands and crosslinks distributed in the network. (a) Stress-stretch curve. (b) The fraction of broken strands and the fraction of broken crosslinks as a function of stretch. (c) Snapshots of the sample at several stretches. The location of the broken bonds up to (d) $\lambda = 5$ and (e) $\lambda = 8$, where they are color-coded according to the stretch at which a bond breaks. \label{fig:reg_hist}}
\end{figure}

Define stress, S, by the force in the current state divided by the cross-sectional area in the undeformed state. Define stretch, $\lambda$, by the current length of the simulation cell in the stretching direction divided by the length of the undeformed simulation cell. The stress-stretch curve predicted by the simulation goes up, peaks, and declines (Fig.~2(a)). We identify the strength by the peak stress, which is $\sim21$ MPa. This strength is $\sim200$ times lower than the strength of our quartic bond, which is $4.4$ GPa. \response{The numerical ratio between peak stress and bond-strength scale does not depend on the specific coarse-grained bond potential and scission criterion, as it scales the peak stress and bond-strength by the same amount.} This prediction of a large reduction in strength is consistent with numerous experimental results. For example, a representative value of the experimentally measured strength of natural rubber is $\sim20$ MPa~\citep{wang2023research}, which is $\sim500$ times lower than the strength of C-C covalent bond ($\sim10$ GPa)~\citep{yang2019polyacrylamide}. Throughout the stretching process, only a small fraction of strands and crosslinks break (Fig.~\ref{fig:reg_hist_b}).  Even at the peak stress, which is attained at a stretch of $\lambda = 5$, less than $5\%$ of the strands and less than $2\%$ of the crosslinks in the network break.
\subsection{Non-local Bond Breaking}
At several stretches, we plot all beads, colored by the original chain to which they belong (Fig.~\ref{fig:reg_hist_c}). Holes much larger than individual beads form at a large stretch of $\lambda = 8$. But no such holes form at the peak stress ($\lambda = 5$). As the network is stretched, we record the stretch and location at which each individual bond breaks. We divide the coordinates of each broken bond by the dimensions of the simulation cell at the moment of bond breaking, and plot the scaled coordinates as points, color-coded by the stretch at which the bond breaks. Up to a stretch of  $\lambda = 5$, when the stress peaks, the bonds break throughout the network, not in any localized region (Fig.~\ref{fig:reg_hist_d}). Up to a stretch of $\lambda = 8$, when the stress has already declined significantly, there is still not a strong indication of bond breaking disproportionally in certain localized regions (Fig.~\ref{fig:reg_hist_e}). A clearer indication of localized bond breaking is only observed at even larger stretches such as  $\lambda = 10$ (Supplementary Fig.~\ref{supp_fig:bond_map}). These observations support the conclusion that the existence of a crack is not necessary for the large reduction in strength. 

\begin{figure}[H]
    \centering
    \includegraphics[width=0.9\linewidth]{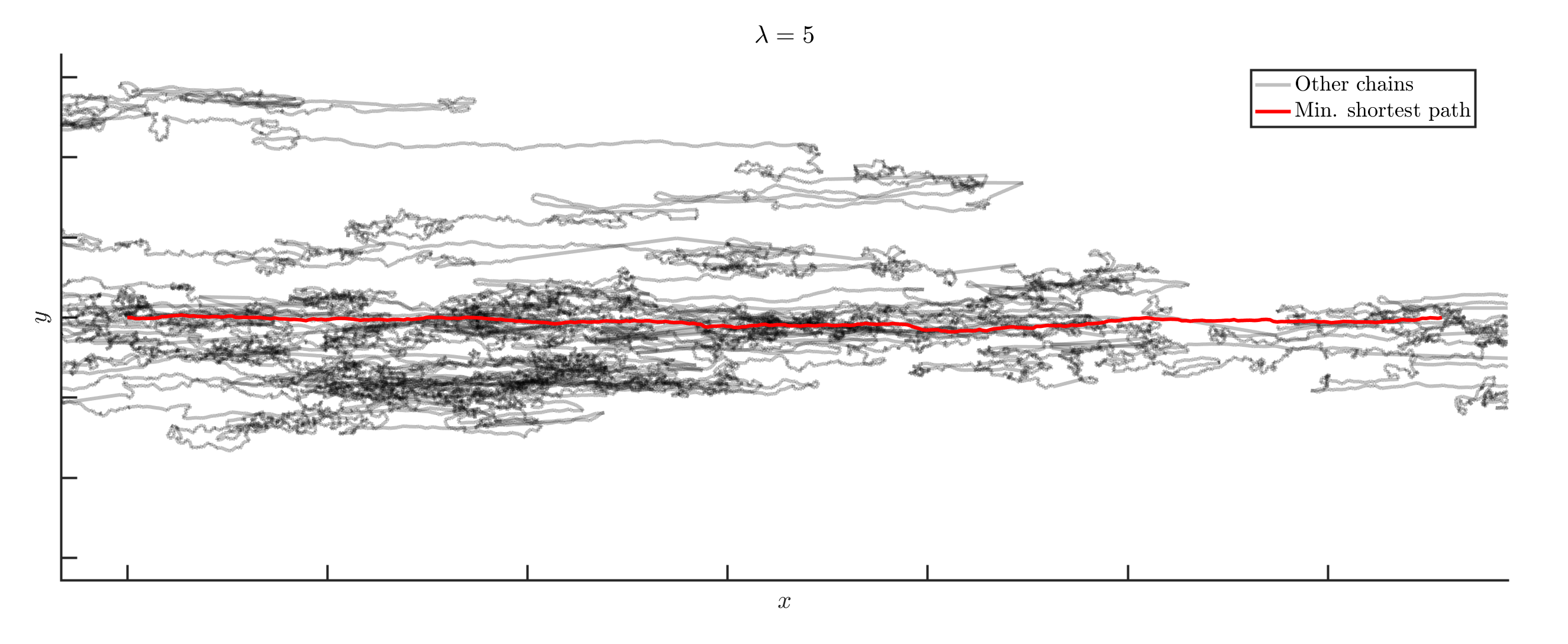}
    \caption{Snapshots of a network at $\lambda = 5$, the minimum shortest path is marked red, and strands connected to it are marked gray. The minimum shortest path is stretched nearly straight, near the breaking point, but the other strands deform by entropic elasticity unless they themselves lie on the shortest paths that are similar in length to the minimum shortest path.}
    \label{fig:side_tofu}
\end{figure}

\change{To visualize the shape of shortest paths belonging to the left-tail of the length distribution, we} plot the shape of the minimum shortest path (in red) together with several other strands at the stretch $\lambda = 5$, where the stress peaks (Fig.~\ref{fig:side_tofu}).  The minimum shortest path is nearly straight, while most of the strands in the network are not unless they happen to lie on a \change{left-tail shortest path} which is similar in length to the minimum shortest path. These relatively straighter paths and segments comprise a small fraction of the network and bear the high tension approaching the covalent bond strength, while a large fraction of the paths bear low stress. These limited high-tension paths contribute negligibly to overall load-bearing, while the majority of low-tension paths bear most of the load. This network ``imbalance'' causes the polymer to rupture at a stress that is orders of magnitude below the strength of the covalent bonds.

\section{Discussions} \label{sec:disc}
\begin{figure}[H]
    \centering
    \includegraphics[width=0.48\linewidth]{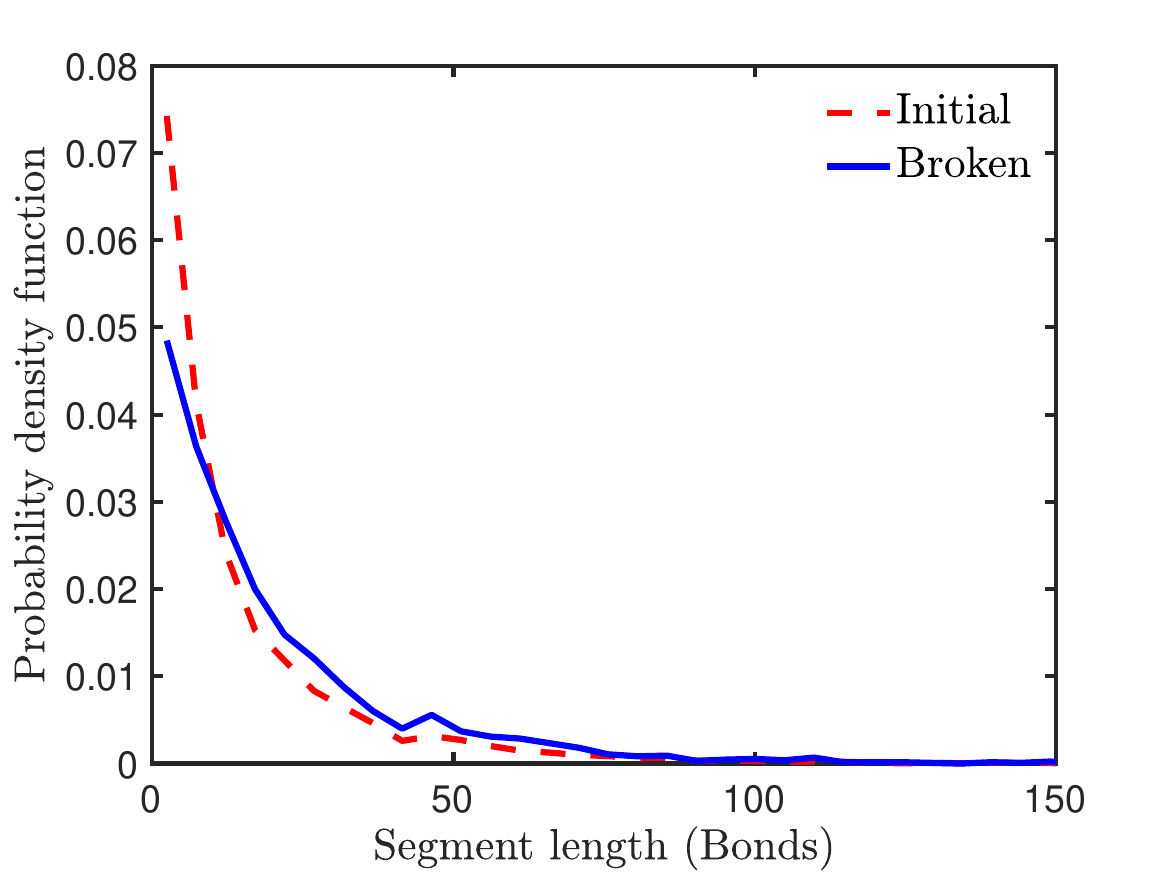}
    \caption{Distributions of the strand length of the undeformed configuration (red dashed curve) and of all the broken strands (blue solid curve). Here the length of a strand is measured by the number of bonds in the strand.}
    \label{fig:comp_pdf_fraction}
\end{figure}

\response{The present work differs in scope and analysis from prior shortest-path studies. In contrast to substrate-constrained adhesive layers \citep{stevens2001interfacial,stevens2001manipulating}, the simulation cell is subjected to periodic boundary conditions in all three directions to represent a bulk network devoid of any edge/interface effects. In contrast to studies focused primarily on predicting the onset (first) scission events \citep{yu2025shortest}, our emphasis is on the macroscopic strength reduction (peak stress) and on how this reduction arises from (i) sequential breaking of only a small fraction of bonds prior to the peak stress and (ii) the evolution of shortest-path length distribution during loading and damage.} It has been hypothesized that bond breaks preferentially on short strands~\citep{tehrani2017effect,edgecombe1998role,hill2023relationship} but this hypothesis contradicts our simulation. We plot the distribution of lengths of all strands in the undeformed network, and the distribution of lengths of all broken strands in the network up to stretch $\lambda = 11$ (Fig.~\ref{fig:comp_pdf_fraction}). These two distributions nearly coincide, indicating that short strands are not more likely to break than long strands.

\begin{figure}[H]
    \centering
    \subfigure[]{\includegraphics[width=0.48\textwidth]{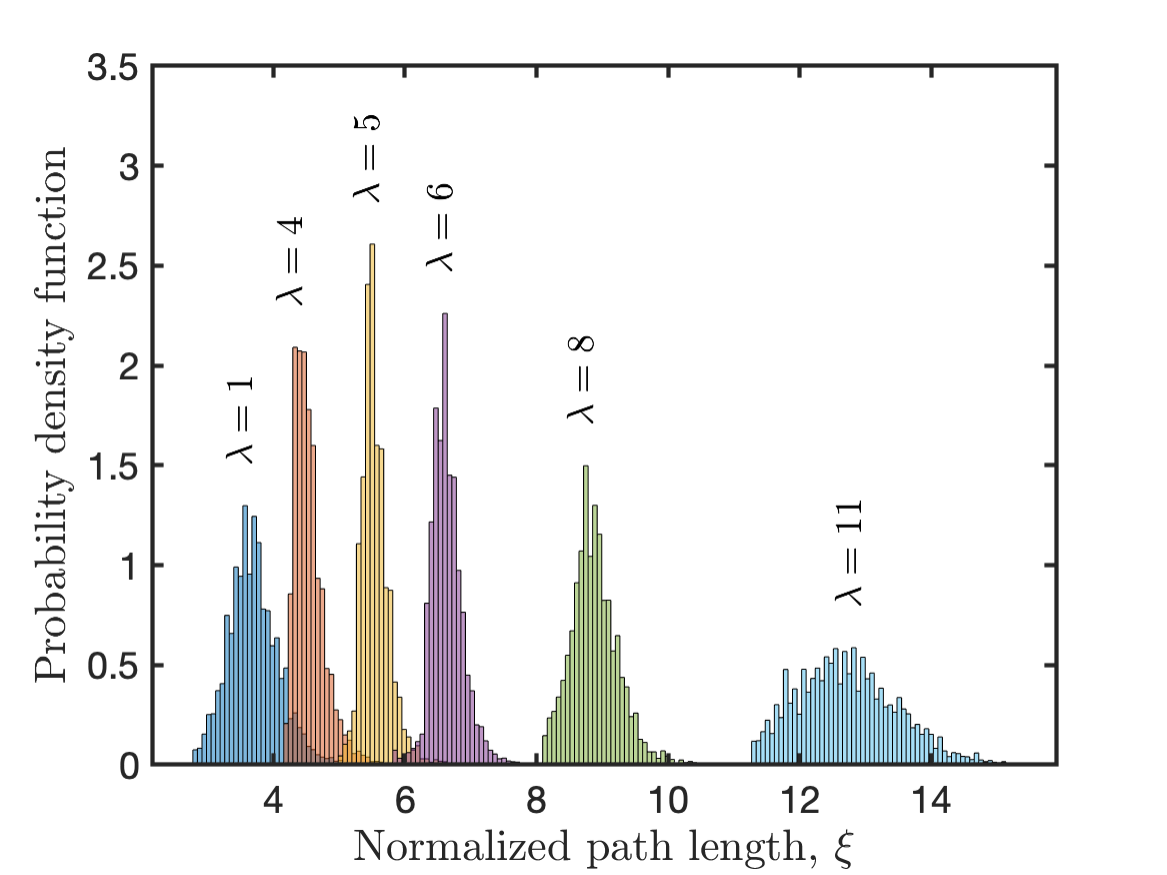}\label{fig:reg_dist_a}}
   \subfigure[]{\includegraphics[width=0.48\textwidth]{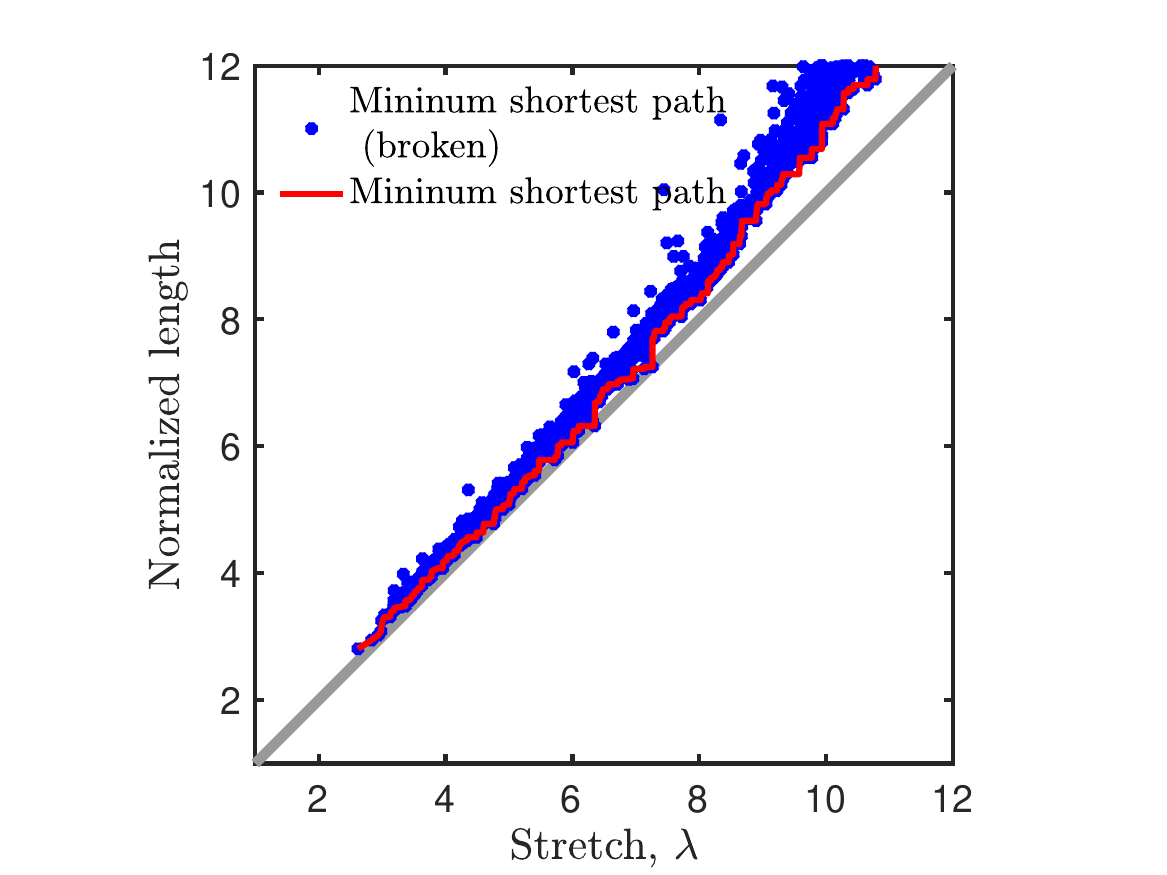}\label{fig:reg_dist_b}}
    \caption{As a network is stretched, the distribution of the shortest path length evolves. (a) Distribution of the normalized shortest path length at several stretches. (b) The minimum among all shortest path lengths as a function of stretch (red solid line). The minimum shortest path length among the broken shortest paths as a function of stretch (blue dots). A $45^{\circ}$ line (gray) is shown to guide the eye. }
    \label{fig:reg_dist}
\end{figure}

We next report the evidence that a network ruptures by breaking bonds on a sequence of \change{left-tail} shortest paths. At each stretch $\lambda$, we count the number of bonds for each shortest path between every pair of nodes that are separated in the direction of stretch by the simulation cell size. Define the normalized path length, $\xi$, by the number of the bonds of a path divided by ($L_0/b$), where $L_0$ is the initial length of the simulation cell, and $b$ is the bond length. For each stretch, we plot the distribution of the normalized shortest path length (Fig.~\ref{fig:reg_dist_a}). Observe that the left edge of each distribution is the minimum shortest path, which coincides with the stretch when the stretch is sufficiently large, $\lambda > 3$. 

As the network is stretched and bonds are broken, the scatter in normal lengths of the shortest paths first narrows, and then broadens (Fig.~\ref{fig:reg_dist_a}). 
%
\response{Recall that all these shortest paths have the same end-to-end distance ($L_0 \lambda$).
The shortest paths with longer contour lengths $\xi$ (i.e. containing more bonds along the paths) are subjected to lower tension.}
%
\response{This statement reflects a statistical consequence of chain elasticity. 
%
Paths with more bonds (larger $\xi$) can still accommodate deformation entropically and carry a lower tension force. In contrast, the shortest (left-tail) paths approach full extension and thus carry the highest tension.}
\response{Therefore, a} narrow distribution of $\xi$ means more shortest paths contribute significantly to the overall load. Conversely, a wide distribution means fewer paths contribute significantly to carrying the load. 
This coincides with the initial rise of stress with stretch (up to $\lambda = 5$) and the subsequent decrease of stress with stretch (Fig.~\ref{fig:reg_hist_a}). 

We further plot the minimum among all $\xi$ as a function of stretch (red curve) (Fig.~\ref{fig:reg_dist_b}). Observe that this curve nearly coincides with the diagonal of the figure, indicating that the minimum or the shortest $\xi$ coincides with the stretch (when $\lambda > 3$). Also plotted is the minimum $\xi$ among the broken shortest paths as a function of stretch (blue dots). Observe that these dots \change{lie close to} the red curve for most of the stretches, indicating that bonds almost always break \change{on the minimum shortest path} \change{or on one of the left-tail shortest paths}. This finding confirms what we have noted before: when the network is stretched, the length of the simulation cell increases, and approaches the length of the minimum shortest path, straightening it, and breaking a bond \change{on a left-tail shortest path} \response{(see Supplementary Fig.~\ref{fig:placeholder_sp_tail_dist})}. Furthermore, \change{out of all} the shortest paths, only a small fraction of \change{them} are very close to being straight, and thus only a small fraction of these shortest paths carry a relatively high load. This is because these paths exhibit an inverse Langevin~\citep{kuhn1942beziehungen,zhang2000single} force response\change{, which states that as a path becomes nearly straight, the effective force increases rapidly \response{(see Supplementary Fig.~\ref{fig:placeholder_straight})}. As a result, only those shortest paths whose straightness (end-to-end distance over contour length) exceeds about $0.97$ experience this rapid force increase and are considered high load-carrying \response{(see Supplementary Fig.~\ref{fig:placeholder_straight_thresh})}.}
\response{A highly simplified model, in which every shortest path in the SP length distribution is assumed to carry a force given by the inverse Langevin function, is sufficient to qualitatively capture the appearance of a maximum stress as a function of applied stretch and a strength within a factor of 2 of the CGMD predictions
(see Supplementary Figs.~\ref{supp_fig:force_dist} and \ref{fig:placeholder_stress}).}
This further supports our reasoning behind the rupture of a polymer at a stress that is orders of magnitude below the strength of the covalent bonds. \response{Sequential failure leading to macroscopic rupture after only a small fraction of elements break has also been reported in fibrous biomaterial networks such as fibrin clots, analyzed through multiscale damage mechanics \citep{kliuchnikov2025strength}.}

The bead-spring model reproduces the experimental observation that the strength of a network is orders of magnitude lower than the strength of covalent bonds. The simulation shows that this enormous reduction in strength results from a simple fact: the network ruptures by a sequence of rare events–breaking a small fraction of bonds. Each broken bond lies on 
\change{one of the left-tail shortest paths}, which straightens and bears high tension set by covalent bonds, while most strands off of 
\change{these left-tail shortest paths} are more coiled and carry low tension. After a bond on 
\change{these paths} breaks, the process repeats for 
\change{one of the strands on the surviving left-tail shortest paths}. As the network is stretched and bonds are broken, the scatter in lengths of the shortest paths first narrows, causing stress to rise, and then broadens, causing stress to decline. This sequential breaking of a small fraction of bonds that lie on nearly straight shortest paths causes the network to rupture at a stress that is orders of magnitude below the strength of the covalent bonds. So long as a network is concerned, bonds do not break at a crack tip, but at wherever the sequence of \change{left-tail} shortest paths is. The fundamental understanding obtained from this work can lead to rational strategies to improve the strength of polymer networks by engineering their microstructure.

\section*{Data Availability}
All data can be obtained on request from the corresponding author. 
\section*{Code Availability}
The code for the preparation of the polymer network and the subsequent stretching simulations is available as an open-source code and can be accessed from
\url{https://gitlab.com/micronano_public/polymer_md}.
The code for the analytic predictions of the shortest path is available as open-source code and can be accessed from 
\url{https://gitlab.com/micronano_public/PolyBranchX}.
\section*{Acknowledgements}
S.M. and W.C. acknowledge support from the Precourt Pioneering Project of Stanford University. S.M., J.B., Z.S., and W.C. acknowledge support from the Air Force Office of Scientific Research under award number FA9550-20-1-0397. WC acknowledges support from National Science Foundation, United States under Award Number DMREF 2118522.
\change{The authors would like to thank Myung Chul Kim for his efforts in refining the numerical results.
}
\section*{Author Contributions}
S.M., J.B., Z.S., and W.C. conceived the research and contributed to the writing of the manuscript. W.C. and S.M.  designed the research. S.M. conducted the research, prepared the datasets and figures, and developed the computational code used to generate the results presented in the manuscript.
\section*{Conflict of Interest}
The authors declare no competing financial or non-financial interests.
\section*{Supplementary Information}
Can be accessed \href{https://static-content.springer.com/esm/art%3A10.1038%2Fs41524-026-02143-5/MediaObjects/41524_2026_2143_MOESM1_ESM.pdf}{here}.


\begin{thebibliography}{29}
\providecommand{\natexlab}[1]{#1}
\providecommand{\url}[1]{\texttt{#1}}
\expandafter\ifx\csname urlstyle\endcsname\relax
  \providecommand{\doi}[1]{doi: #1}\else
  \providecommand{\doi}{doi: \begingroup \urlstyle{rm}\Url}\fi

\bibitem[Yang et~al.(2019)Yang, Yin, and Suo]{yang2019polyacrylamide}
Canhui Yang, Tenghao Yin, and Zhigang Suo.
\newblock Polyacrylamide hydrogels. i. network imperfection.
\newblock \emph{Journal of the Mechanics and Physics of Solids}, 131:\penalty0
  43--55, 2019.

\bibitem[Helaly et~al.(2011)Helaly, El~Sabbagh, El~Kinawy, and
  El~Sawy]{helaly2011effect}
FM~Helaly, SH~El~Sabbagh, OS~El~Kinawy, and SM~El~Sawy.
\newblock Effect of synthesized zinc stearate on the properties of natural
  rubber vulcanizates in the absence and presence of some fillers.
\newblock \emph{Materials \& Design}, 32\penalty0 (5):\penalty0 2835--2843,
  2011.

\bibitem[Wang et~al.(2023)Wang, Liao, Wang, Yu, Zheng, Lian, Luo, Liao, Liu,
  and Peng]{wang2023research}
Yueqiong Wang, Lusheng Liao, Rui Wang, Heping Yu, Tingting Zheng, Yujia Lian,
  Mingchao Luo, Shuangquan Liao, Hongchao Liu, and Zheng Peng.
\newblock Research of strain induced crystallization and tensile properties of
  vulcanized natural rubber based on crosslink densities.
\newblock \emph{Industrial Crops and Products}, 202:\penalty0 117070, 2023.

\bibitem[Clarke and Faber(1987)]{clarke1987fracture}
DR~Clarke and KT~Faber.
\newblock Fracture of ceramics and glasses.
\newblock \emph{Journal of Physics and Chemistry of Solids}, 48\penalty0
  (11):\penalty0 1115--1157, 1987.

\bibitem[Shand(1965)]{shand1965strength}
Errol~B Shand.
\newblock Strength of glass—the griffith method revised.
\newblock \emph{Journal of the American Ceramic Society}, 48\penalty0
  (1):\penalty0 43--49, 1965.

\bibitem[C{\'a}ceres et~al.(1995)C{\'a}ceres, Davidson, and
  Griffiths]{caceres1995deformation}
CH~C{\'a}ceres, CJ~Davidson, and JR~Griffiths.
\newblock The deformation and fracture behaviour of an al7 si0.4 mg casting
  alloy.
\newblock \emph{Materials Science and Engineering: A}, 197\penalty0
  (2):\penalty0 171--179, 1995.

\bibitem[Berry(1961)]{berry1961fracture}
JP~Berry.
\newblock Fracture processes in polymeric materials. ii. the tensile strength
  of polystyrene.
\newblock \emph{Journal of Polymer Science}, 50\penalty0 (154):\penalty0
  313--321, 1961.

\bibitem[Kendall et~al.(1983)Kendall, Howard, and
  Birchall]{kendall1983relation}
K~Kendall, AJ~Howard, and James~Derek Birchall.
\newblock The relation between porosity, microstructure and strength, and the
  approach to advanced cement-based materials.
\newblock \emph{Philosophical Transactions of the Royal Society of London.
  Series A, Mathematical and Physical Sciences}, 310\penalty0 (1511):\penalty0
  139--153, 1983.

\bibitem[Yin et~al.(2020)Yin, Bertin, Wang, Bao, and Cai]{yin2020topological}
Yikai Yin, Nicolas Bertin, Yanming Wang, Zhenan Bao, and Wei Cai.
\newblock Topological origin of strain induced damage of multi-network
  elastomers by bond breaking.
\newblock \emph{Extreme Mechanics Letters}, 40:\penalty0 100883, 2020.

\bibitem[Yin et~al.(2024)Yin, Mohanty, Cooper, Bao, and Cai]{yin2024network}
Yikai Yin, Shaswat Mohanty, Christopher~B Cooper, Zhenan Bao, and Wei Cai.
\newblock Network evolution controlling strain-induced damage and self-healing
  of elastomers with dynamic bonds.
\newblock \emph{arXiv preprint arXiv:2401.11087}, 2024.

\bibitem[Zhang et~al.(2024)Zhang, Mohanty, Blanchet, and
  Cai]{zhang2024modeling}
Zhenyuan Zhang, Shaswat Mohanty, Jose Blanchet, and Wei Cai.
\newblock Modeling shortest paths in polymeric networks using spatial branching
  processes.
\newblock \emph{Journal of the Mechanics and Physics of Solids}, 187:\penalty0
  105636, 2024.

\bibitem[Mohanty(2024)]{mohanty2024understanding}
Shaswat Mohanty.
\newblock \emph{Understanding the Microstructural and Macroscopic Evolution of
  Dynamic Polymer Networks through Coarse-grained Molecular Dynamics}.
\newblock PhD thesis, Stanford University, 2024.
\newblock URL \url{https://purl.stanford.edu/vv258vb0486}.

\bibitem[Tehrani and Sarvestani(2017)]{tehrani2017effect}
Mohammad Tehrani and Alireza Sarvestani.
\newblock Effect of chain length distribution on mechanical behavior of
  polymeric networks.
\newblock \emph{European Polymer Journal}, 87:\penalty0 136--146, 2017.

\bibitem[Edgecombe et~al.(1998)Edgecombe, Stein, Fr{\'e}chet, Xu, and
  Kramer]{edgecombe1998role}
Brian~D Edgecombe, Jason~A Stein, Jean~MJ Fr{\'e}chet, Zhihua Xu, and Edward~J
  Kramer.
\newblock The role of polymer architecture in strengthening polymer- polymer
  interfaces: a comparison of graft, block, and random copolymers containing
  hydrogen-bonding moieties.
\newblock \emph{Macromolecules}, 31\penalty0 (4):\penalty0 1292--1304, 1998.

\bibitem[Tao et~al.(2023)Tao, Lavoie, Suo, and Cameron]{tao2023effect}
Manyuan Tao, Shawn Lavoie, Zhigang Suo, and Maria~K Cameron.
\newblock The effect of scatter of polymer chain length on strength.
\newblock \emph{Extreme Mechanics Letters}, 61:\penalty0 102024, 2023.

\bibitem[Yu and Jackson(2025)]{yu2025shortest}
Zheng Yu and Nicholas~E Jackson.
\newblock Shortest paths govern bond rupture in thermoset networks.
\newblock \emph{Macromolecules}, 58\penalty0 (3):\penalty0 1728--1736, 2025.

\bibitem[Stevens(2001{\natexlab{a}})]{stevens2001interfacial}
Mark~J Stevens.
\newblock Interfacial fracture between highly cross-linked polymer networks and
  a solid surface: effect of interfacial bond density.
\newblock \emph{Macromolecules}, 34\penalty0 (8):\penalty0 2710--2718,
  2001{\natexlab{a}}.

\bibitem[Stevens(2001{\natexlab{b}})]{stevens2001manipulating}
Mark~J Stevens.
\newblock Manipulating connectivity to control fracture in network polymer
  adhesives.
\newblock \emph{Macromolecules}, 34\penalty0 (5):\penalty0 1411--1415,
  2001{\natexlab{b}}.

\bibitem[Kremer and Grest(1990)]{kremer1990dynamics}
Kurt Kremer and Gary~S Grest.
\newblock Dynamics of entangled linear polymer melts: A molecular-dynamics
  simulation.
\newblock \emph{The Journal of Chemical Physics}, 92\penalty0 (8):\penalty0
  5057--5086, 1990.

\bibitem[Thompson et~al.(2022)Thompson, Aktulga, Berger, Bolintineanu, Brown,
  Crozier, in~'t Veld, Kohlmeyer, Moore, Nguyen, Shan, Stevens, Tranchida,
  Trott, and Plimpton]{LAMMPS}
A.~P. Thompson, H.~M. Aktulga, R.~Berger, D.~S. Bolintineanu, W.~M. Brown,
  P.~S. Crozier, P.~J. in~'t Veld, A.~Kohlmeyer, S.~G. Moore, T.~D. Nguyen,
  R.~Shan, M.~J. Stevens, J.~Tranchida, C.~Trott, and S.~J. Plimpton.
\newblock {LAMMPS} - a flexible simulation tool for particle-based materials
  modeling at the atomic, meso, and continuum scales.
\newblock \emph{Comp. Phys. Comm.}, 271:\penalty0 108171, 2022.

\bibitem[Sliozberg and Andzelm(2012)]{sliozberg2012fast}
Yelena~R Sliozberg and Jan~W Andzelm.
\newblock Fast protocol for equilibration of entangled and branched polymer
  chains.
\newblock \emph{Chemical Physics Letters}, 523:\penalty0 139--143, 2012.

\bibitem[Uddin and Ju(2016)]{uddin2016multiscale}
Md~Salah Uddin and Jaehyung Ju.
\newblock Multiscale modeling of a natural rubber: Bridging a coarse-grained
  molecular model to the rubber network theory.
\newblock \emph{Polymer}, 101:\penalty0 34--47, 2016.

\bibitem[Chaikumpollert et~al.(2012)Chaikumpollert, Yamamoto, Suchiva, and
  Kawahara]{chaikumpollert2012mechanical}
Oraphin Chaikumpollert, Yoshimasa Yamamoto, Krisda Suchiva, and Seiichi
  Kawahara.
\newblock Mechanical properties and cross-linking structure of cross-linked
  natural rubber.
\newblock \emph{Polymer journal}, 44\penalty0 (8):\penalty0 772--777, 2012.

\bibitem[Ge et~al.(2013)Ge, Pierce, Perahia, Grest, and
  Robbins]{ge2013molecular}
Ting Ge, Flint Pierce, Dvora Perahia, Gary~S Grest, and Mark~O Robbins.
\newblock Molecular dynamics simulations of polymer welding: Strength from
  interfacial entanglements.
\newblock \emph{Physical review letters}, 110\penalty0 (9):\penalty0 098301,
  2013.

\bibitem[Dijkstra(2022)]{dijkstra2022note}
Edsger~W Dijkstra.
\newblock A note on two problems in connexion with graphs.
\newblock In \emph{Edsger Wybe Dijkstra: His Life, Work, and Legacy}, pages
  287--290. ACM Books, 2022.

\bibitem[Hill and Ronan(2023)]{hill2023relationship}
Aoife Hill and William Ronan.
\newblock Relationship between failure strain, molecular weight, and chain
  extensibility in biodegradable polymers.
\newblock \emph{Journal of the Mechanical Behavior of Biomedical Materials},
  139:\penalty0 105663, 2023.

\bibitem[Kuhn and Gr{\"u}n(1942)]{kuhn1942beziehungen}
Werner Kuhn and Felix Gr{\"u}n.
\newblock Beziehungen zwischen elastischen konstanten und
  dehnungsdoppelbrechung hochelastischer stoffe.
\newblock \emph{Kolloid-Zeitschrift}, 101\penalty0 (3):\penalty0 248--271,
  1942.

\bibitem[Zhang et~al.(2000)Zhang, Zou, Wang, and Zhang]{zhang2000single}
Wenke Zhang, Shan Zou, Chi Wang, and Xi~Zhang.
\newblock Single polymer chain elongation of poly (n-isopropylacrylamide) and
  poly (acrylamide) by atomic force microscopy.
\newblock \emph{The Journal of Physical Chemistry B}, 104\penalty0
  (44):\penalty0 10258--10264, 2000.

\bibitem[Kliuchnikov et~al.(2025)Kliuchnikov, Dagklis, Litvinov, Marx, Weisel,
  Bassani, Purohit, and Barsegov]{kliuchnikov2025strength}
Evgenii Kliuchnikov, Angelos~Gkarsen Dagklis, Rustem~I Litvinov, Kenneth~A
  Marx, John~W Weisel, John~L Bassani, Prashant~K Purohit, and Valeri Barsegov.
\newblock Strength, deformability, damage and fracture toughness of fibrous
  material networks: Application to fibrin clots.
\newblock \emph{Acta Biomaterialia}, 201:\penalty0 347--359, 2025.

\end{thebibliography}
\end{document}